\documentclass[12pt]{iopart}
\usepackage{amstext}
\usepackage{setstack}
\usepackage{graphicx}

\newcommand{\ket}[1]{|#1\rangle}
\newcommand{\bra}[1]{\langle#1|}
\newcommand{\vev}[1]{\langle#1\rangle}
\newcommand{\inner}[2]{\langle#1|#2\rangle}

\newcommand{\norm}[1]{\|#1\|}
\newcommand{\bpsil}{\bra{\psi_L}}
\newcommand{\bpsir}{\bra{\psi_R}}
\newcommand{\kpsil}{\ket{\psi_L}}
\newcommand{\kpsir}{\ket{\psi_R}}
\newcommand{\rhosys}{\rho_{\text{sys}}}
\newcommand{\Trenv}{\Tr_{\text{env}}}

\begin{document}

\title[Density matrix in the stochastic TMRG]{On the choice of the
  density matrix\\in the stochastic TMRG}
\author{T~Enss and U~Schollw\"ock}
\address{Sektion Physik, Ludwig-Maximilians-Universit\"at M\"unchen,
  Theresienstr.~37/III, D-80333 M\"unchen, Germany}
\eads{\mailto{Tilman.Enss@physik.uni-muenchen.de}, 
  \mailto{scholl@theorie.physik.uni-muenchen.de}}
\begin{abstract}
  In applications of the density matrix renormalization group to
  non\-hermitean problems, the choice of the density matrix is not
  uniquely prescribed by the algorithm. We demonstrate that for the
  recently introduced stochastic transfer matrix DMRG (stochastic
  TMRG) the necessity to use open boundary conditions makes
  asymmetrical reduced density matrices, as used for renormalization
  in quantum TMRG, an inappropriate choice.  An explicit construction
  of the largest left and right eigenvectors of the full transfer
  matrix allows us to show why symmetrical density matrices are the
  correct physical choice.
\end{abstract}
\pacs{02.50.Ey, 64.60.Ht, 02.70.-c, 05.10.Cc}
\submitto{\JPA}
\date{June 11, 2001}


\section{Introduction}

Since its inception in 1992 by White\cite{White:1992}, the Density
Matrix Renormalization Group (DMRG) has emerged as one of the most
powerful numerical methods in the study of low dimensional strongly
correlated fermionic, bosonic or quantum magnetic
systems\cite{Peschel:1999}.  The universality of its core idea of
deducing a decimation prescription for state spaces by considering the
renormalization flow of suitable reduced density matrices has led many
researchers to extend the method to other fields of research in
systems with many correlated degrees of freedom.

Originally, the method was applied to the renormalization of effective
low-energy Hamiltonians to study static and dynamic $T=0$ properties.
Major progress occurred with Nishino's realisation\cite{Nishino:1995}
that the DMRG can be used to renormalize the transfer matrix of
semi-infinite two-dimensional strips, a method which we will refer to
as TMRG (transfer matrix renormalization group) in the following.  The
use of a transfer matrix implies that the system is truly infinite in
one spatial direction, whereas the power of the DMRG to treat large
one-dimensional systems is reflected in the fact that the strip width
can be chosen so large as to allow reliable finite-size
extrapolations.  In analogy to the approach used in Quantum Monte
Carlo, one-dimensional quantum systems were then mapped by a
Trotter-Suzuki checkerboard decomposition to a two-dimensional
classical system\cite{Bursill:1996}.  The renormalization of the
resulting quantum transfer matrix allows the study of the
thermodynamics of quantum chains for infinite system sizes and very
fine Trotter-Suzuki decompositions, i.e.\ down to very low
temperatures.

As the Trotter-Suzuki decomposition generates a transfer matrix which
is not invariant under a spatial reflection of the system, the quantum
transfer matrix is asymmetrical and there are left and right
eigenstates $\bpsil$ and $\kpsir$ associated with the
thermodynamically relevant largest eigenvalue of the transfer matrix.
This raises the question which is the correct way to construct the
optimal density matrix for the DMRG.  In the usual hermitean DMRG, the
density matrix is typically obtained by tracing out a part of the
universe (``environment'') in the ground state:
\begin{equation}
  \rhosys = \Trenv \ket{\psi}\bra{\psi} .
\end{equation}
In the nonhermitean case, Bursill
\etal\cite{Bursill:1996} made the {\em symmetrical} choice
\begin{equation}
  \rhosys = \Trenv \kpsir\bpsir
\end{equation}
which can be easily extended to 
\begin{equation}
  \rhosys^{\text{symm}} = \Trenv (\kpsir\bpsir + \ket{\psi_L}\bpsil)
\end{equation}
which treats the left and right eigenvectors on an equal footing (all
composite density matrices have to be appropriately normalized,
depending on the normalization chosen for the eigenvectors; we do not
show these factors explicitly in the paper). This approach also has
the property to minimize the sum of the two squared distances between
the original eigenstates and their projections onto the reduced state
spaces after the renormalization step, which is the prescription from
which the DMRG procedure for symmetrical matrices can be derived.

However, it was pointed out by Wang and Xiang\cite{Wang:1997} that the
choice of an {\em asymmetrical} density matrix,
\begin{equation}
  \rhosys^{\text{asym}} = \Trenv \kpsir\bpsil,
\end{equation}
is more physically adequate and yields numerically more satisfying
results.  This is now the accepted choice for the application of the
TMRG to quantum
problems\cite{Shibata:1997,Maisinger:1998,Nishino:1999}.

A further field of applications was opened up by the application of
the DMRG idea to the renormalization of genuinely asymmetrical
problems\cite{Hieida:1998,Kaulke:1998,Carlon:1999} as they occur in
nonequilibrium statistical mechanics.  Carlon, Henkel and
Schollw\"{o}ck\cite{Carlon:1999} exploited the fact that the time
evolution of reaction-diffusion systems can be mapped to a
Schr\"{o}dinger-like equation.  In this case, one can study the
long-time behaviour of finite size systems quite
precisely\cite{Carlon:1999,Carlon:2001,Carlon:2001a,Henkel:2001}.

Here, the question of the correct choice of the density matrix arises
as the transition matrix is genuinely asymmetrical itself, since there
is no detailed balance. Carlon \etal\ found the choice of the
symmetrical density matrix to be most suitable, but essentially due to
reasons of numerical stability: in this class of problems, very high
numerical precision for the chosen basis states in the reduced basis
was found to be important, which is less easily obtained for
asymmetrical matrices. An attempt to include projectors like $\kpsir
\bpsil$ while building a symmetrical density matrix by studying
\begin{equation}
  \rhosys^{\text{mix}} = \Trenv (\kpsil + \kpsir) (\bpsil + \bpsir),
\end{equation} 
yielded inferior results.

Recently, the TMRG has been used by Kemper \etal\ to successfully
renormalize stochastic transfer matrices\cite{Kemper:2001}, an
approach which henceforth we will refer to as ``stochastic TMRG''. In
their study of the Domany-Kinzel cellular automaton, they found that
the use of a symmetrical density matrix yielded satisfactory results,
while the asymmetrical choice led to incorrect results, which is at
variance with the conventional (or ``quantum'') TMRG. These findings
were obtained on an entirely empirical basis.

In this paper, we want to investigate in depth the question of the
correct choice of the density matrix, which is at the core of all
asymmetrical DMRG applications, for the new stochastic TMRG. We will
give explicit constructions for the eigenvalues and selected
eigenvectors of the stochastic transfer matrices, highlighting the
central role of the boundary conditions in time direction. This will
allow us to demonstrate that the content of the asymmetrical density
matrix is essentially trivial, while the symmetrical density matrix is
the physically adequate choice.


\section{Explicit construction of the unrenormalized TMRG transfer matrix}
\label{sec:TMRG}

\subsection{The TMRG transfer matrix}
\label{sec:TMRG:def}

\begin{figure}[ht]
  \begin{center}
    \includegraphics{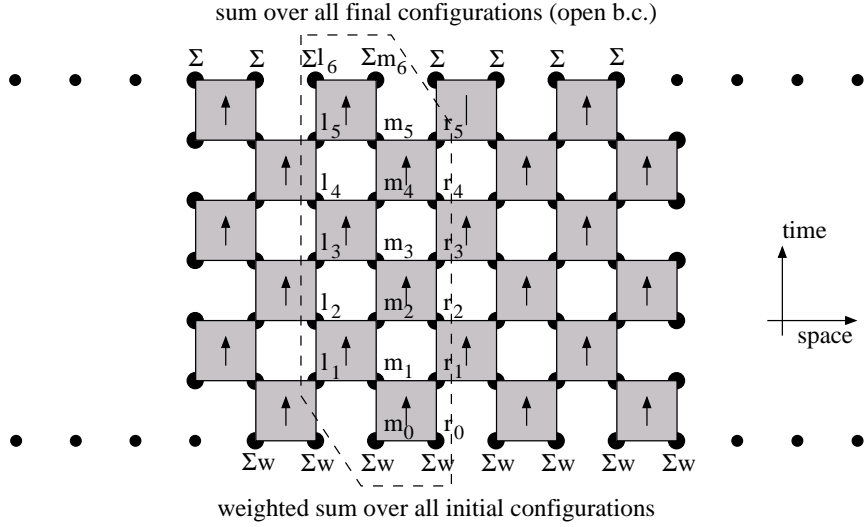}
  \end{center}
  \caption{\label{fig:trotter}Trotter-Suzuki decomposition of the full
  time evolution into local interactions $\tau$ ($2M=6$ time steps
  shown).  The lattice is spatially infinitely extended, while in time
  direction the system is finite.  Two adjacent columns (boxed)
  constitute the basic building block: the transfer matrix
  $(T)^{l_1\dots l_{2M-1}}_{r_1\dots r_{2M-1}}$.}
\end{figure}

In the stochastic TMRG (for a description of the very similar TMRG
applied to quantum systems, see \cite{Peschel:1999}), one considers a
stochastically evolving system extended infinitely in one spatial
dimension with (for simplicity) local update rules involving
neighbouring sites only that allow a finite number of states ($n$).
The local interaction between two (or similarly, a few) lattice sites
is given by the local transfer matrix
\begin{equation}
  (\tau)_{l_1r_1}^{l_2r_2}
  = [\exp(\Delta t\cdot H_{\text{local}})]_{l_1r_1}^{l_2r_2}
  = \raisebox{-0.5cm}{\includegraphics{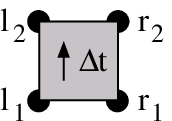}}
\end{equation}
where the stochastic ``Hamiltonian'' $H_{\text{local}}$ gives the
transition rates between states and the arrow indicates the direction
of the time step $\Delta t$.  We assume spatial parity invariance
\begin{equation}
  \label{eq:parity}
  (\tau)_{l_1r_1}^{l_2r_2} =
  (\tau)_{r_1l_1}^{r_2l_2}\,.
\end{equation}
In complete analogy to Quantum Monte Carlo and the conventional TMRG,
the real time evolution operator for the full lattice is mapped
approximately by a Trotter-Suzuki decomposition to a checkerboard
structure of local interactions (\fref{fig:trotter}).

In the time direction the evolution of the system is simulated during
a finite interval.  As illustrated in \fref{fig:trotter}, the full
evolution operator is then written as $\exp(2M\Delta t\cdot H) =
\lim_{N\to\infty} T^N$, where $M$ is the Trotter number, $N$ the
(diverging) system size and $H = \lim_{N\rightarrow\infty}
\sum_{i=-N}^N H_{\text{local}}(i,i+1)$.  $T$ is called the transfer
matrix which consists of two adjacent columns of $\tau$'s, where we
take as upper (lower) index the ``history'' (time evolution) of the
lattice site at the left (right) side of the zig-zagged column:
\begin{equation}
  \fl (T)^{l_1\dots l_{2M-1}}_{r_1\dots r_{2M-1}}
  = \underbrace{\sum_{m_0r_0}w(m_0)\,w(r_0)}_{\text{open initial b.c.}}
  \sum_{{m_1\dots\\ m_{2M-1}}}
  \underbrace{\sum_{l_{2M}m_{2M}}}_
  {\makebox[1mm][c]{\scriptsize open final b.c.}}
  \prod_{k=0}^{M-1} (\tau)_{m_{2k}r_{2k}}^{m_{2k+1}r_{2k+1}}
  (\tau)_{l_{2k+1}m_{2k+1}}^{l_{2k+2}m_{2k+2}}
\end{equation}
where $l$, $m$, and $r$ denote the left, middle and right lattice
sites of $T$, respectively.  One uses stochastic initial conditions:
the weight of each possible initial configuration is given by the
product of local weights $w(s_i)$ of the states $s_i$ at each lattice
site $i$,
\begin{equation}
  \label{eq:normw}
  w(\dots,s_1,s_2,s_3,\dots) = \prod_i w(s_i),
  \qquad \sum_{s_i=1}^{n} w(s_i) = 1 \text{ (norm.)}
\end{equation}
such that if each state is weighted equally, there is no bias in the
initial configuration.  At the end, one uses open final boundary
conditions, i.e.\ all final states are allowed without bias.  In the
diagrams, we thus implicitly trace over all internal indices, sum over
all final indices, and perform a weighted sum over all initial
indices.

\subsection{Spectrum and eigenvectors of the transfer matrix}
\label{sec:TMRG:spec}

\begin{figure}[ht]
  \begin{center}
    \includegraphics{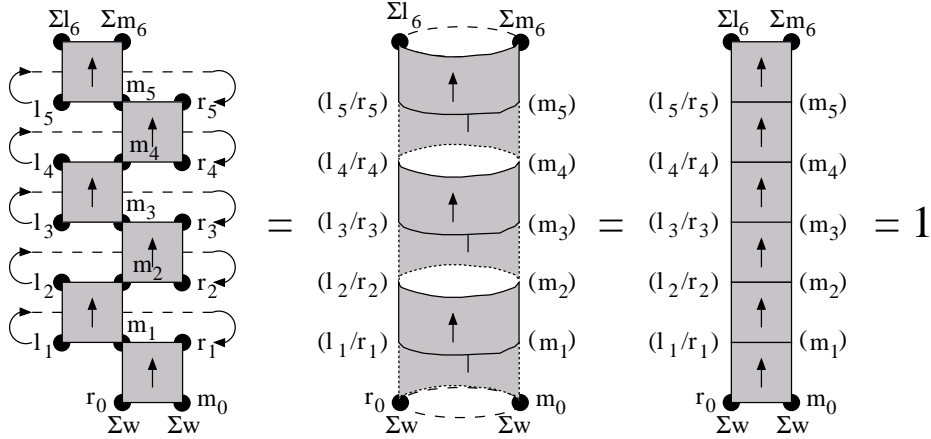}
  \end{center}
  \caption{\label{fig:cylinder}$\Tr (T)_{r_1\dots r_{2M-1}}^{l_1\dots
      l_{2M-1}}$ is equivalent to $T$ being rolled up on a cylinder
    with a solely local interaction.}
\end{figure}

As the local stochastic transfer matrix $\tau$ must obey probability
conservation, the probability of ending up in \emph{any} state is
unity:
\begin{equation}
  \label{eq:sumtau}
  \sum_{l_2r_2} (\tau)_{l_1r_1}^{l_2r_2} = 1 \quad
  \forall l_1,r_1, \qquad
  \raisebox{-0.5cm}{\includegraphics{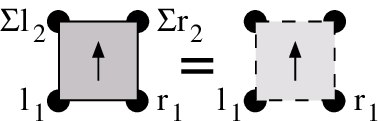}} \text{(trivial)}.
\end{equation}
This implies
\begin{equation}
  \label{eq:traceT}
  \Tr T=1 ,
\end{equation}
where tracing contracts the upper index of $T$ with the lower one;
thus the probability that the lattice sites at the left and right
boundary of $T$ have the same time evolution is unity. In pictorial
language, the trace rolls up the transfer matrix into a cylinder, with
its left and right boundary identified and the middle site at the
opposite side of the cylinder (cf.\ \fref{fig:cylinder}).

But this is equivalent to having only the local interaction between
the left/right lattice site on one side and the middle one on the
other, for $2M$ time steps $\Delta t$, which is known exactly (e.g.\ 
$2M=6$):
\begin{equation}
  \fl (T)_{r_1\dots r_5}^{l_1\dots l_5} = 
  \sum_{m_0r_0} w(m_0) \,w(r_0) \sum_{l_6m_6} \sum_{m_1\dots m_5} 
  (\tau)_{l_5m_5}^{l_6m_6} (\tau)_{m_4r_4}^{m_5r_5} \cdots
  (\tau)_{l_1m_1}^{l_2m_2} (\tau)_{m_0r_0}^{m_1r_1}  
\end{equation}
\begin{eqnarray}
  \Tr T & = \sum_{l_1\dots l_5} (T)_{l_1\dots
    l_5}^{l_1\dots l_5} \nonumber\\
  & = \sum_{m_0r_0} w(m_0)\, w(r_0) \sum_{l_6m_6}
  \sum_{l_1\dots l_5} \sum_{m_1\dots m_5} 
  (\tau)_{l_5m_5}^{l_6m_6} (\tau)_{m_4l_4}^{m_5l_5} \cdots
  (\tau)_{m_0l_0}^{m_1l_1} \nonumber\\
  & = \sum_{m_0r_0} w(m_0)\, w(r_0) \sum_{l_6m_6}
  \left[ \tau^{2M}(\Delta t) \right]_{r_0m_0}^{l_6m_6}
  \quad \text{(by parity invar.\ (\ref{eq:parity}))} \nonumber\\
  & = \sum_{m_0r_0} w(m_0)\, w(r_0) \sum_{l_6m_6}
  \left[ \exp(2M\Delta t\cdot H_{\text{local}})
  \right]_{r_0m_0}^{l_6m_6} \nonumber\\
  & = \sum_{m_0r_0} w(m_0)\, w(r_0) \sum_{l_6m_6}
  \bigl[ \tau(2M\Delta t) \bigr]_{r_0m_0}^{l_6m_6} \nonumber\\
  & = \sum_{m_0r_0} w(m_0)\, w(r_0) \cdot 1 = 1 \quad
  \text{(by (\ref{eq:sumtau}) and (\ref{eq:normw}))}. \nonumber
\end{eqnarray}

Next we consider the spectrum and eigenvector decomposition of $T$.
As all entries of $T$ are non-negative, the Frobenius theorem implies
that the largest eigenvalue $\lambda_0$ is non-degenerate and real.
The associated left and right eigenvectors $\bpsil$ and $\kpsir$
(where $\kpsir^\dagger \neq \bpsil$ in general, since $T$ is not
symmetrical) have nonnegative entries.  Now the decisive difference
between conventional and stochastic TMRG is that in the conventional
TMRG one has periodic boundary conditions in (imaginary) time, but
open final boundary conditions in the stochastic TMRG.  This latter
property implies
\begin{equation}
  \label{eq:decomp}
  T^{2M-1} = \kpsir \bpsil = T^k \quad \forall k\ge 2M-1\,,
\end{equation}
i.e.\ sufficiently large powers of $T$ are identical and decompose
into an outer product of the two eigenvectors of $\lambda_0=1$,
normalized as $\inner{\psi_L}{\psi_R} = 1$, while all other
eigenvalues of $T$ vanish.

\begin{figure}[ht]
  \begin{center}
    \includegraphics{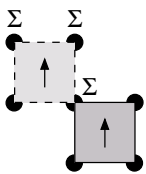}
  \end{center}
  \caption{\label{fig:trivtau}Open b.c.\ lead to trivial $\tau$'s that
    cause also the final indices of the previous $\tau$'s to be summed
    over, thus ``propagating'' the open b.c.\ backwards in time.}
\end{figure}

\begin{figure}[ht]
  \begin{center}
    \includegraphics{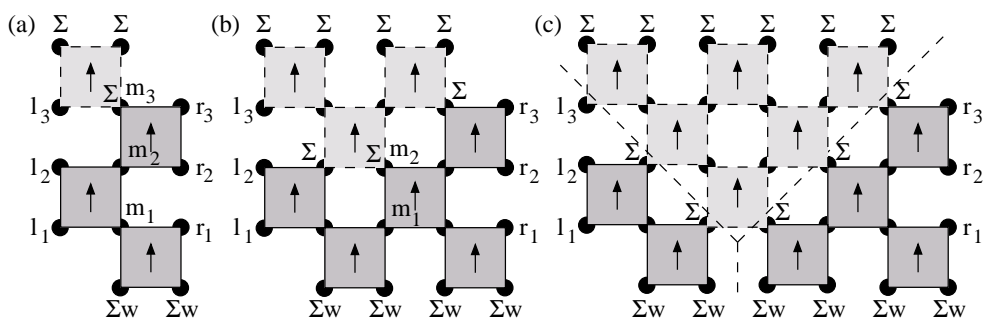}
  \end{center}
  \caption{\label{fig:taupow}$T^k$ decompose into a left and right part
    of ever smaller rank until $T^{2M-1}$ decomposes into two
    lightcones which cannot interact during $2M$ time steps (here
    $2M=4$): (a) $T = (\psi_R)_{m_1m_2}^{l_1l_2l_3}\cdot
    (\psi_L)_{r_1r_2r_3}^{m_1m_2}$ with 2 connecting indices
    $(m_1,m_2)$, (b) $T^2 = (\psi_R)_{m_1}^{l_1l_2l_3}\cdot
    (\psi_L)_{r_1r_2r_3}^{m_1}$ with one connecting index $(m_1)$, (c)
    $T^{2M-1} = (\psi_R)^{l_1l_2l_3}\cdot (\psi_L)_{r_1r_2r_3} =
    \kpsir\bpsil$ with no connecting index.}
\end{figure}

To see this, consider \fref{fig:trivtau}.  Because $\tau$ is
stochastic, it becomes trivial by summing over its final indices (cf.\ 
(\ref{eq:sumtau}); its value does not depend on its initial indices,
such that the contraction with the final index of a \emph{previous}
$\tau$ yields an unweighted sum over the previous $\tau$'s final
index.  Therefore in \fref{fig:taupow}, the trivial $\tau$'s
``propagate'' the open final b.c.\ backwards in time, resulting in a
cone of trivial $\tau$'s that have no influence on the value of the
transfer matrix $T^k$.

The Trotter-Suzuki decomposition introduces a finite speed of action
(``speed of light'') into the model: during each time step, only
nearest neighbours that are connected by an edge of a
(\emph{nontrivial}) $\tau$ can interact, so interactions can only
propagate at a speed of one lattice site per time step $\Delta t$.
Hence we speak of the ``light cone'' of a certain site as those sites
in time and space that influence the probabilities of different
occupations for this site.

In $T$ (cf.\ \fref{fig:taupow}a), the last $\tau$ is trivial, hence it
has the value 1 indepedent of $l_3$ and $m_3$.  Thus we effectively
sum over the internal index $m_3$, and $l_3$ is a ``dummy index'': the
value of $(T)_{r_1r_2r_3}^{l_1l_2l_3}$ does not depend on it.  This
leaves only two indices $(m_1,m_2)$ connecting the left and right part
of $T$, so the transfer matrix can be decomposed into a product of two
matrices $\psi_R$ and $\psi_L$ of rank $n^2$ ($n$ is the number of
possible states per site).

When we multiply this with another $T$ (cf.\ \fref{fig:taupow}b), we
observe that a third $\tau$ becomes trivial as the final $\tau$'s
propagate the final open b.c.\ backwards.  This leads to a sum also
over $m_2$, leaving $m_1$ as the only connecting index, so $T^2$
decomposes into a product of two matrices of rank $n$.

Finally in $T^{2M-1}$ (cf.\ \fref{fig:taupow}c), the trivial $\tau$'s
propagate all the way to the initial $\tau$'s, effectively separating
the left and right boundary: now there are no connecting indices left
between $\psi_R$ and $\psi_L$, turning them into column and row
vectors $\kpsir$ and $\bpsil$.  From the viewpoint of causality, the
left and right boundary are separated by $4M-2$ lattice sites, so no
initial lattice site can influence \emph{both} the left and the right
boundary, i.e.\ the light cone for the lattice sites on the left
boundary ($l_1,\dots,l_{2M-1}$) does not intersect the light cone for
the lattice sites on the right boundary ($r_1,\dots,r_{2M-1}$).  This
means that the probability of a particular history on the left
boundary is completely independent of the probability of a particular
history on the right.

We can now explicitly compute these column and row vectors
$\kpsir^{l_1\dots l_{2M-1}}$ and $\bpsil_{r_1\dots r_{2M-1}}$: we take
the {\em left} light cone (of triangular shape, delimited by the
dashed line) and contract the included nontrivial local transfer
matrices $\tau$ in all internal, initial, and final indices, keeping
the \emph{free} indices $l_1\dots l_{2M-1}$, and denote this object as
({\em right} eigenvector, see below) $\kpsir^{l_1\dots l_{2M-1}}$.
Analogously, we denote the same construction on the right with free
indices $r_1\dots r_{2M-1}$ as $\bpsil_{r_1\dots r_{2M-1}}$.  Note
that the last index on the left, $l_{2M-1}$, is by construction a
dummy index which does not change the value of $\kpsir^{l_1\dots
  l_{2M-1}}$.

If we consider even higher powers of $T$ than $2M-1$, we only add more
trivial $\tau$'s in the middle that lie outside either light cone and
therefore do not change the value of the transfer matrix.  Thus we
have proven $T^{2M-1} = T^k\;\forall k\ge 2M-1$, which implies for the
eigenvalues $\lambda_i$ of $T$ that $\lambda_i^{2M-1} =
\lambda_i^{2M}\;\forall i$, such that $\lambda_i$ can only assume the
values 0 or 1.  From (\ref{eq:traceT}) follows $\lambda_0=1$,
$\lambda_i=0\; \forall i>0$.

\begin{figure}[ht]
  \begin{center}
    \includegraphics{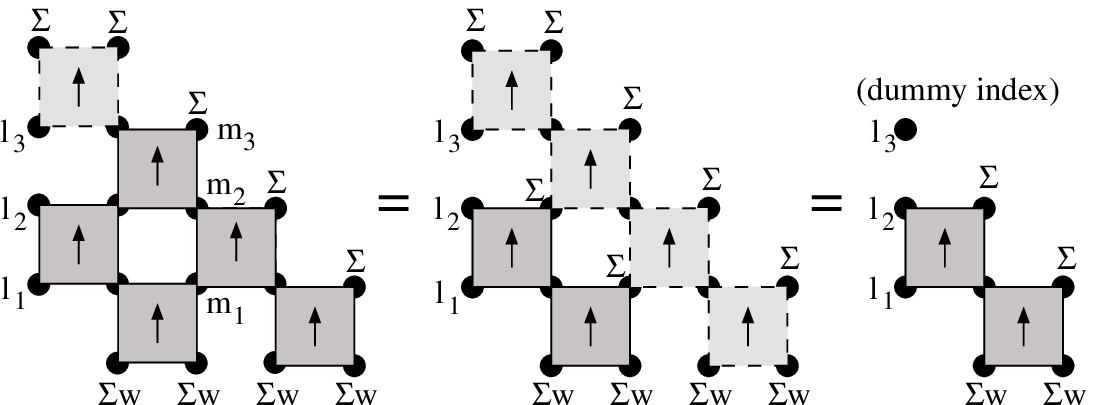}
  \end{center}
  \caption{\label{fig:eigenvec}$\kpsir$ is indeed the right eigenvector
    of $T$ corresponding to the eigenvalue $\lambda_0=1$: for $2M=4$,
    $T_{m_1m_2m_3}^{l_1l_2l_3} \ket{\psi_R}^{m_1m_2m_3} =
    \ket{\psi_R}^{l_1l_2l_3}$.}
\end{figure}

The above decomposition of $T^{2M-1}$ into an outer product of two
vectors $\kpsir\bpsil$ is in fact the explicit eigenvector
decomposition corresponding to the eigenvalue $\lambda_0=1$:
multiplying the right (left) eigenvector onto $T^{2M-1}$ from the
right (left) is, in pictorial language, attaching the respective light
cone to the transfer matrix with a trace over all overlapping indices.
But now the final open b.c.\ can propagate through all $\tau$'s on
that side, turning them trivial (which automatically gives the correct
normalization $\inner{\psi_L}{\psi_R} = 1$) and leaving just the light
cone on the other side, which reproduces exactly the previously
attached eigenvector.  The eigenvalue 1 is determined by this
normalization condition:
\begin{equation}
  \label{eq:eigenvec}
  T^{2M-1}\kpsir = \kpsir \inner{\psi_L}{\psi_R}
  = \kpsir \cdot 1\,.
\end{equation}
This concludes the proof of (\ref{eq:decomp}).  Similarly, we check
that $\kpsir$ and $\bpsil$ are also the eigenvectors of $T$
corresponding to the same eigenvalue (cf.\ \fref{fig:eigenvec}).

Let us finally emphasize that for small powers of $T$, $T^k$ with
$k<2M-1$, this simple decomposition does not hold. But this is not
relevant, as we take $k$ to infinity in the TMRG.

To arrive at a prescription for the stochastic TMRG density matrix, we
must now consider how actual physical information is extracted from
the stochastic transfer matrix. This is done by computing expectation
values of local operators, averaged over the whole lattice at final
time $2M\cdot \Delta t$ by multiplying $\tau$ in the last time step
with the local operator, e.g.\ for the local particle number operator
$n(i)$ one has
\begin{equation}
  (\tau_n)_{l_1r_1}^{l_2r_2} :=
  (\tau)_{l_1r_1}^{l_2r_2} \cdot n(l_2)
\end{equation}
where we denote the transfer matrix with its last $\tau$ replaced by
$\tau_n$ as $T_n$ (coloured black in the diagram).  Then the
expectation value of the particle number operator can be calculated as
\begin{eqnarray}
  \makebox[0cm][r]{$\vev{n_{t=2M\cdot\Delta t}}$}
  & = \lim_{N\to\infty} \frac{\Tr (T^N T_n)} {\Tr (T^{N+1})} 
    = \lim_{N\to\infty} \frac{\Tr \left( \raisebox{-6mm}
          {\includegraphics{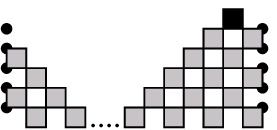}} \right)}
      {\Tr \left( \raisebox{-5mm}{\includegraphics{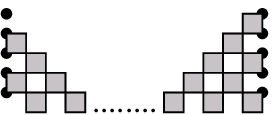}} \right)} \\
  & = \frac{\Tr (\kpsir \bpsil T_n)} {\Tr (\kpsir \bpsil)} 
    = \frac{\bpsil T_n \kpsir}
      {\underbrace{\inner{\psi_L}{\psi_R}}_{1}}
    = \left( \, \raisebox{-6mm}{\includegraphics{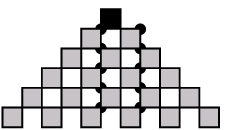}} \, \right)
\end{eqnarray}
where we have omitted local transfer matrices that become trivial by
summing and used the pictorial construction of the eigenvectors.  The
computation of expectation values is thus reduced to the computation
of the light cone of lattice sites that can possibly influence the
particular site at which the operator is measured.

\begin{figure}[ht]
  \begin{center}
    \includegraphics[scale=0.93]{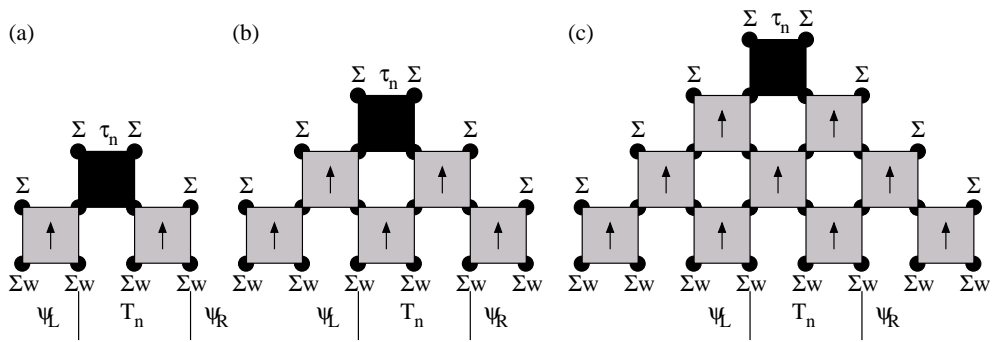}
  \end{center}
  \caption{\label{fig:lightcones}The light cones that are needed to compute
    the expectation value of the particle number operator for the
    first few time steps: (a) $2M=2$, (b) $2M=3$, (c) $2M=4$.}
\end{figure}

\Fref{fig:lightcones} shows the light cones for the first few time
steps.  In this manner one can compute the exact evolution of
$\vev{n(t)}$, the only inaccuracy stemming from the Trotter-Suzuki
decomposition. It is interesting to observe that for determining both
eigenvectors and expectation values, a matrix diagonalization can be
completely avoided.

For the choice of the density matrix the important observation in
\fref{fig:lightcones} is now that by parity invariance
(\ref{eq:parity}) and adding one dummy index ($i_{2M-1}$),
\begin{equation}
  \label{eq:psilr}
  \kpsir_{2M}^{i_1\dots i_{2M-1}}
  = ( \bpsil_{2M-1,\,i_1\dots i_{(2M-1)-1}} )^T
  =: \ket{\psi_L}_{2M-1}^{i_1\dots i_{2M-2}}
\end{equation}
i.e.\ \emph{the left eigenvector for a given time step is identical to
  the right eigenvector in the next (half) time step.}

\section{Density matrix renormalization of the stochastic
transfer matrix}
\label{sec:TMRG:renorm}

We have seen how, in principle, physical information can be extracted
from the stochastic transfer matrix, but both memory usage and
computation time grow exponentially with the number of time steps.
The transfer matrix operates on the space of possible time evolutions
of particular sites, which has to be decimated to remain numerically
manageable. This is the idea of the TMRG.  As the probability of a
partial time evolution is determined not only by its past but also by
its future continuation, and the infinite future is not yet known, one
contends oneself with a ``lookahead'' of several time steps which is
called the ``environment'', while the past time evolution --- that is
being projected onto a smaller Hilbert space --- is called the
``system''.

\begin{figure}[ht]
  \begin{center}
    \includegraphics[scale=0.86]{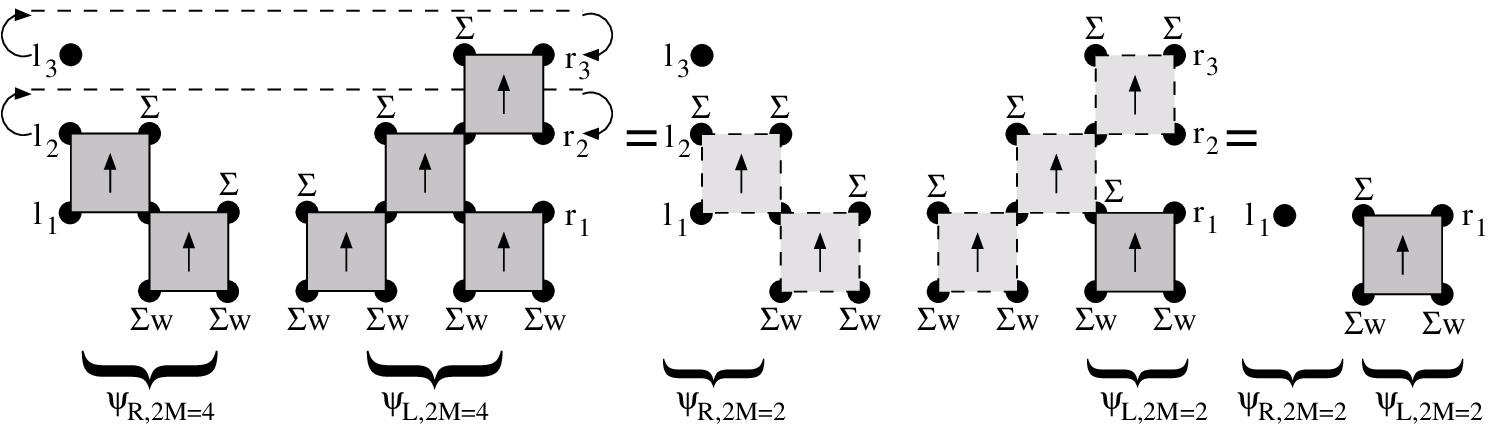}
  \end{center}
  \caption{\label{fig:trenv}Computing the asymmetrical density matrix
    $\rhosys^{\text{asym}}$: the trace makes the environment trivial,
    $\Trenv \kpsir \bpsil = \kpsir_{\text{sys}} \bpsil_{\text{sys}}$.}
\end{figure}

Traditionally, one obtains this basis by computing the reduced density
matrix for the system, $\rhosys$, by taking a partial trace over the
environment of the outer product of right and left $T$ eigenvectors;
the new Hilbert space basis consists of $\rhosys$'s eigenvectors
associated to the largest eigenvalues (highest probabilities).
Inspired from traditional TMRG (with periodic b.c.\ in time
direction), one could try $\rhosys^{\text{asym}} = \Trenv \kpsir
\bpsil$.  But in the case of open b.c., this is just a partial trace
of the transfer matrix,
\begin{equation}
  \rhosys^{\text{asym}} 
  = \Trenv \kpsir \bpsil
  = \Trenv T^{2M-1}
  = \kpsir_{\text{sys}} \bpsil_{\text{sys}}
\end{equation}
having a spectrum of
\begin{equation}
  \label{eq:specasym}
  \text{Spec}(\rhosys^{\text{asym}}) = \{ 1,0,\dots,0 \}.
\end{equation}
We demonstrate this for $2M=4$, taking e.g.\ the first index as the
system index and the following two indices as environment indices.
Then the trace over the environment $\Trenv = \sum_{l_3r_3l_2r_2}
\delta_{l_3}^{r_3} \delta_{l_2}^{r_2}$, as illustrated in
\fref{fig:trenv}, makes all $\tau$'s in the environment trivial: when
we contract the dummy index $l_3$ with $r_3$, the final $\tau$ of
$\bpsil$ becomes trivial, so $r_2$ now becomes a dummy index.  This in
turn gives a sum over $l_2$, making also the final $\tau$ of $\kpsir$
trivial, etc.  In the end we are left with $\kpsir$ and $\bpsil$ of an
earlier time step, namely the one with only the system indices $l_1$,
$r_1$.

Thus this asymmetrical density matrix provides only a single basis
vector for the Hilbert space projection, which is obviously not
enough.  The reason is that the open b.c.\ ``average away'' the
physical interaction in the environment by making it irrelevant how
the system will react in the future. In these boundary conditions
resides the essential difference between quantum and stochastic TMRG.

Each renormalization step using $\rhosys^{\text{asym}}$ projects
therefore onto a one-dimensional Hilbert space.  No longer-ranged
correlation implicit in differently weighted basis vectors is being
built up.  We therefore expect to observe only the local evolution of
the model, which is exactly what we got from numerical simulation
using the asymmetrical density matrix.

\begin{figure}[ht]
  \begin{center}
    \includegraphics{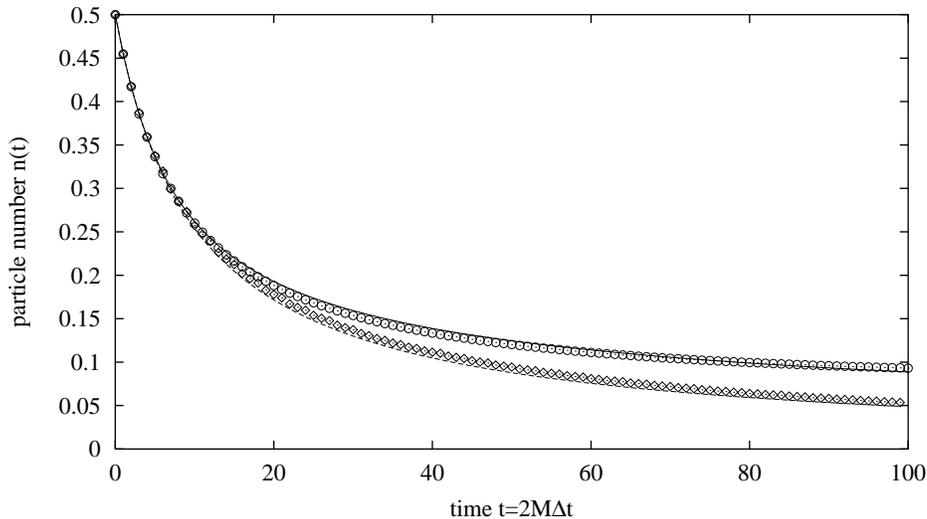}
  \end{center}
  \caption{\label{fig:reacdiff}Comparing the simulation of a
    reaction-diffusion model for symme\-trized vs.\ asymmetrical density
    matrices.  (\full) exact solution, (\dashed) exact local (4-site)
    interaction, (\opencircle) simulation, symmetrized density matrix,
    (\opendiamond) simulation, asymmetrical density matrix ($D=0.1$ in
    inverse time units, and $\Delta t=1$).}
\end{figure}

As an example, we use an exactly solvable reaction-diffusion model:
$AA\rightarrow 00$ with rate $2\alpha$ (reaction) and
$A0\leftrightarrow 0A$ with rate $D$ (diffusion).  For $2\alpha=D$,
the exact solution is given by Spouge\cite{Spouge:1988}: for initial
partical number $n(t=0)=\case{1}{2}$ (unbiased $w(0) = w(A) =
\case{1}{2}$, i.e.\ all lattice sites have an independent probability
$\case{1}{2}$ of occupancy),
\begin{equation}
  \label{eq:rhoex}
  \vev{n(t)} = \case{1}{2} \exp(-2Dt) \left[ I_0(2Dt)+I_1(2Dt) \right]
  \sim \frac{1}{\sqrt{t}} \quad \text{(asympt.)}
\end{equation}
This agrees well only with the simulation using the symmetrized
density matrix $\rhosys^{\text{symm}}$ (cf.\ \fref{fig:reacdiff}).
But when we instead use the asymmetrical density matrix
$\rhosys^{\text{asym}}$, we observe a $1/t$ decay.  This resembles
very closely what we expect from a local interaction: the exact 4-site
interaction (i.e.\ renormalizing as soon as the system has $n$ states)
is
\begin{equation}
  \label{eq:ex4}
  \vev{n(2M+1)} = \vev{n(2M)} - [1-\exp(-D\Delta t)] \cdot
  \vev{n(2M)}^2
\end{equation}
($n(0)=\case{1}{2}$) which goes asymptotically as $1/t$.

\begin{figure}[ht]
  \begin{center}
    \includegraphics{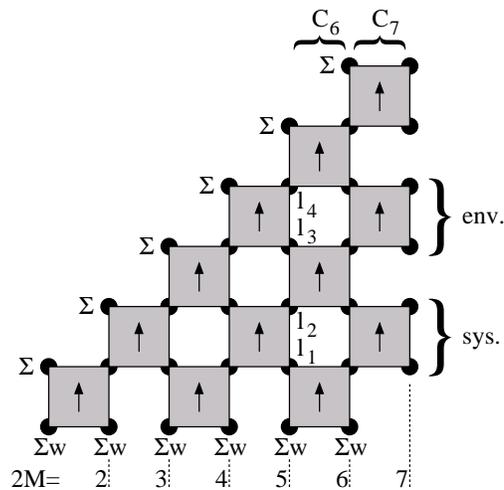}
  \end{center}
  \caption{\label{fig:growpsi}The evolution of the largest left $T$
    eigenvector $\bpsil_{2M}$ for the first few Trotter steps.}
\end{figure}

The origin of this shortcoming becomes clear in the following
consideration: \fref{fig:growpsi} shows the left eigenvectors during
consecutive time steps.  Suppose we want to find a good basis for
renormalizing $\bpsil_{2M=5}$, taking its first two indices ($l_1$,
$l_2$) as system and the latter two ($l_3$, $l_4$) as environment.
After somehow tracing over the environment, the new basis vectors will
have indices $l_1$, $l_2$, and they will roughly resemble $\bpsil_3$,
having a strong correlation between $l_1$ and $l_2$ because they are
connected by an edge of $\tau$.

Now in order to advance one time step from $\bpsil_5$ to $\bpsil_6$,
we see from \fref{fig:growpsi} that we need to multiply $\bpsil_5$ by
the matrix represented by the column $C_6$.  This reveals the problem:
the first two indices of $C_6$ are not connected by an edge of $\tau$
and therefore completely independent.  Thus the renormalization basis
obtained from $\bpsil_5$ with a strong correlation between the first
pair of indices fits very poorly for renormalizing $C_6$. The states
in the basis are essentially orthogonal to those that would describe a
renormalized $C_6$ well. Looking ahead, $C_7$ which would bring us to
$\bpsil_7$ has again a structure similar to $\bpsil_5$ and could be
renormalized well.

This problem would be remedied by adding to the renormalization basis
also basis vectors of a form similar to $\bpsil_4$ (tracing over the
last index as environment) which shows little correlation between the
first two indices.  Noting that $\bpsil_4 = (\kpsir_5)^T$
(\ref{eq:psilr}) we have to mix left and right eigenvectors,
$\bpsil_5$ and $\bpsir_5$, in order to obtain a good basis.  Based on
the original argument by White\cite{White:1992} about the optimal
density matrix in DMRG, Carlon \etal\cite{Carlon:1999} have shown
that in order to express \emph{two} different vectors $\kpsil$,
$\kpsir$ most accurately (in the \emph{same} basis), one has to use an
eigenbasis of $\rhosys = \Trenv (\kpsir\bpsir + \kpsil\bpsil)$.  The
symmetrical form of this density matrix arises because they used as a
measure of distance between exact and approximated vectors the
symmetrical norm $\norm{\kpsir-\tilde{\kpsir}}^2 +
\norm{\kpsil-\tilde{\kpsil}}^2$.  With this same argument, we now
conclude that \emph{this symmetrized density matrix is also optimal in
  the case of explicit eigenvectors}.

This explains why a symmetrized density matrix, e.g.\ 
$\rhosys^{\text{symm}} = \Trenv (\kpsir\bpsir + \kpsil\bpsil
)$ as used by Kemper \etal\cite{Kemper:2001}, works so well.
Indeed, within the precision of the method, the exact result for the
example considered is reproduced when we use this symmetrized reduced
density matrix $\rhosys^{\text{symm}}$.

\begin{figure}[ht]
  \begin{center}
    \includegraphics{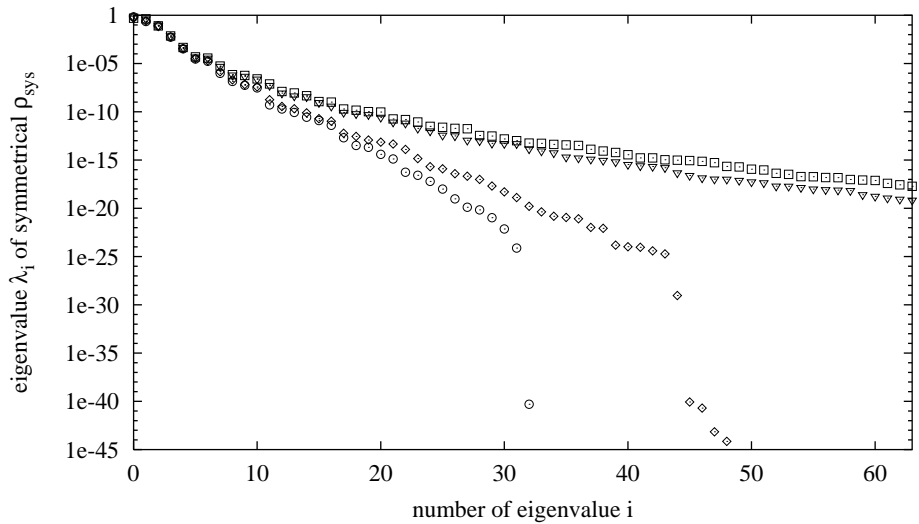}
  \end{center}
  \caption{\label{fig:rhospec}Spectrum of the reduced system density
    matrix $\rhosys^{\text{symm}}$ at specific steps $2M$ ($m=64$).
    (\opencircle) $2M=18$ (trunc.\ $10^{-40}$), (\opendiamond) $2M=20$
    (trunc.\ $10^{-29}$), (\opentriangledown) $2M=60$ (trunc.\ 
    $10^{-21}$), (\opensquare) $2M=280$ (trunc.\ $10^{-21}$).}
\end{figure}

If we consider for our example the eigenvalue spectrum of the
symmetrized density matrix, initially it falls off approximately
exponentially, $\lambda_i \sim \alpha^{-i}$ for some constant $\alpha$
(cf.\ \fref{fig:rhospec}).  After many renormalization steps, the
spectrum falls off rather according to a power law with large negative
exponent, $\lambda_i \sim i^{-\beta}$.  The symmetrized density matrix
therefore indeed fulfills two basic requirements: It should have as
many nonzero eigenvalues as possible in order to keep rich nontrivial
information about the future evolution in the environment (contrary to
(\ref{eq:specasym})), but at the same time they should drop off as
quickly as possible so that when we truncate the spectrum at $m$
states that are retained, we lose as little information on the system
as possible.

Now, in practical applications of the quantum TMRG, system and
environment are treated on an equal basis, renormalizing the system
with a density matrix traced over the environment and vice versa. In
the stochastic TMRG, the explicit construction of $T$ and the
eigenvectors opens the possibility to renormalize only the system and
keep the environment as a ``lookahead'' of constant finite length.
This simplifies the algorithm, since the $\psi$'s can be renormalized
explicitly as shown above, and there is no need to ever solve the
eigenvalue problem of the transfer matrix; on the other hand, due to
the exponential growth of the matrices with a longer lookahead, this
approach is in practice limited (see our numerical example below), and
one reverts to the conventional approach of renormalizing both system
and environment of the transfer matrix $T$ and then finding its
largest eigenvectors anew in each step by using a suitable algorithm
for the diagonalisation of large nonsymmetrical matrices.

Our considerations above were all for the case without renormalized
environment. Renormalizing it, too, leads to an increasingly longer,
but less precise lookahead. As our considerations hold formally for
all lengths of lookaheads, the result on the choice of the density
matrix should hold also in that case, which is indeed what we observe
numerically. To check whether our special choice of a symmetrical
density matrix is indeed optimal among symmetrical density matrices,
we have also considered numerically several other symmetrical density
matrices for comparison, in particular
\begin{eqnarray}
  \label{otherrho}
  \Trenv ( \kpsil\bpsir + \kpsir\bpsil ), \\
  \Trenv ( \kpsil+\kpsir ) ( \bpsil+\bpsir ), \\
  \Trenv ( \kpsil-\kpsir ) ( \bpsil-\bpsir ),
\end{eqnarray}
but the convergence turned out to be worse than for
$\rhosys^{\text{symm}}$. All results shown in the following are
therefore for the ``canonical'' choice of the symmetrical density
matrix. 

\begin{figure}[ht]
  \begin{center}
    \includegraphics{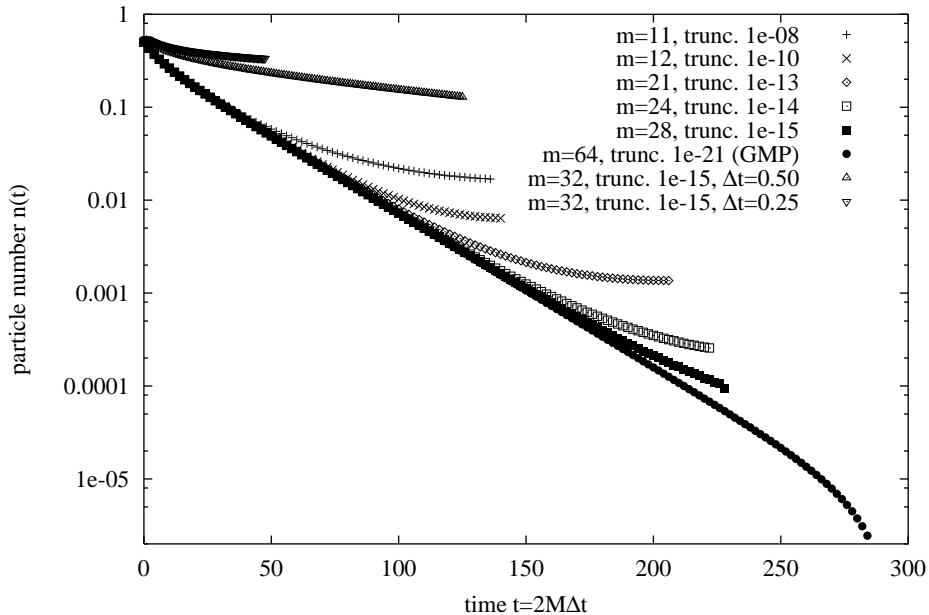}
  \end{center}
  \caption{\label{fig:mmaxeps}Branching-fusing
    model\protect\cite{Carlon:1999} at the critical point
    $p=0.8403578$ for different number of states kept ($m$) resp.\ 
    truncation in the spectrum of $\rhosys$.}
\end{figure}

As an application, we tried the branching-fusing process explained in
\cite{Carlon:1999}: diffusion $A0\leftrightarrow 0A$ (rate $D$), and
several reactions $AA\to 00$ (rate $2\alpha$), $AA\to 0A,A0$ (rate
$\gamma$), $0A,A0\to 00$ (rate $\delta$), and $0A,A0\to AA$ (rate
$\beta$).  The relations between the rates are fixed as $D = 2\alpha =
\gamma = \delta = 1-p\,,\;\; \beta = p$.  There is a critical point of
the directed percolation universality class at $p_c \approx
0.8403578$.  First we check the dependence of our TMRG simulation
(using the conventional algorithm) for different values of $m$ (number
of states kept during projection into a smaller Hilbert space) (cf.\ 
\fref{fig:mmaxeps}) at the critical point, the most difficult point
for the stochastic TMRG.

We observe that the particle number generally decays exponentially,
despite criticality, because we are operating at very short times and
due to the rather rough Trotter decomposition.  On the lower branch,
for a fixed Trotter time $(\Delta t=1)$, our results become more
accurate as we increase $m$, effectively truncating the spectrum of
$\rhosys$ at a smaller eigenvalue, until we reach the machine
precision (53 mantissa bits).  To compute more time steps, one option
is to use more accurate floating point numbers (see below).

For comparison, choosing a smaller time step $\Delta t$ (i.e.\ a finer
Trotter decom\-position, last two curves in \fref{fig:mmaxeps}), while
maintaining a large Hilbert space ($m=32$), we get very different
results: the particle number decreases much more slowly, which shows
that our Trotter decomposition is still far from being sufficiently
fine (of course, extrapolations $\Delta t \rightarrow 0$ are
feasible). Unfortunately the number of time steps we can simulate
decreases, too, which makes such an extrapolation more difficult.
This can be explained as follows: the shorter the time step $\Delta
t$, the smaller the transition probabilities to different states, and
the closer is $\tau$ to the identity matrix.  But a transfer matrix
composed of such $\tau$'s has eigenvectors with exponentially growing
Euclidean norm ($\inner{\psi_R}{\psi_R} \sim n^M$ and analogously for
$\kpsil$).  This is not changed by renormalization because an
orthogonal transformation conserves the norm, and we only lose those
Hilbert space directions that are very improbable.  As the
renormalization keeps the number of components of $\bpsil$ and
$\kpsir$ bounded, the values of the components have to grow
exponentially to obtain a growing norm.  At the same time,
$\inner{\psi_L}{\psi_R} = 1$, so $\bpsil$ and $\kpsir$ are almost
orthogonal.  Now the numerical problem is the following: while the
components grow exponentially, they have to cancel each other very
accurately in order to yield the correct normalization.  This works
only until the components are of order of magnitude
$\sqrt{1/\epsilon}$, where $\epsilon$ is the machine accuracy.
Choosing more accurate floating point numbers, by taking twice as many
mantissa bits we can simulate very roughly twice as many time steps.
This is illustrated in \fref{fig:mmaxeps}: if we use the GNU
multiprecision library (GMP), truncating at $10^{-21}$ instead of
$10^{-15}$, we can simulate approximately 1.5 times as many steps.

\begin{figure}[ht]
  \begin{center}
    \includegraphics{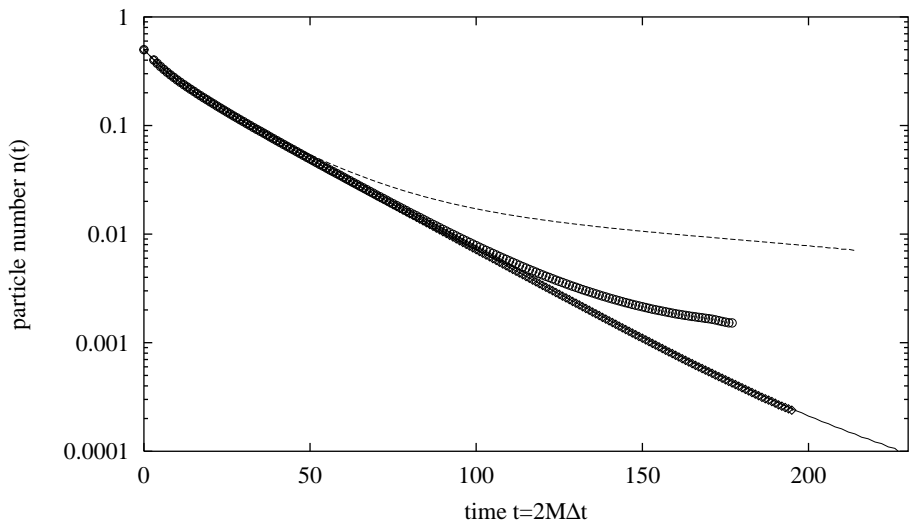}
  \end{center}
  \caption{\label{fig:trialg}Branching-fusing
    model\protect\cite{Carlon:1999} at the critical point
    $p=0.8403578$ for different TMRG methods.  (\full) conventional
    method, $m=28$; (\dashed) conventional method, $m=11$;
    (\opencircle) explicit eigenvectors, $m=8$ (4 sites lookahead);
    (\opendiamond) explicit eigenvectors, $m=11$ (6 sites lookahead).}
\end{figure}

Finally we want to compare the numerical stability of the conventional
TMRG algorithm, renormalizing both system and environment, with the
newly proposed method of explicitly constructing the eigenvectors
(saving the iterations of the transfer matrix diagonalisation which
take up more than half of the time in the original algorithm).  Here
we renormalize only the system, keeping a lookahead of finite length
as environment.  As can be seen in \fref{fig:trialg}, the result is
much better than the conventional algorithm for an intermediate
accuracy of $m=11$ states kept, while it cannot compete with the most
accurate (up to machine precision) values for $m=28$.  So, as already
mentioned above, it is more proof of a concept and a tool for
analytical derivations than for practical simulations.  Neither more
sites lookahead nor larger $m$ improve the result significantly.
The computation time for the displayed curves was in the range of a
few seconds ($m=11$) up to a few tens of seconds ($m=32$) on a PIII
(500 MHz), while memory usage was at most a few MBytes.


\section{Conclusions and Outlook}

We have shown a new way of looking at stochastic TMRG with open b.c.\ 
by explicitly giving the transfer matrix and its left and right
eigenvectors corresponding to the physically relevant eigenvalue 1.
This allowed us to study the physical effect of various choices for
the density matrix of the TMRG. The conclusion was that, unlike the
TMRG for quantum transfer matrices, an unsymmetrical density matrix is
not just less accurate, but fundamentally unsuitable for the
stochastic TMRG. The core problem was that one has open boundary
conditions in the future of the time evolution, averaging away the
physical interaction, which is not the case in the periodically closed
quantum transfer matrix DMRG.  The symmetrical density matrix defined
by demanding an optimal representation of both left and right
eigenvectors in the truncated basis turned out to be the mandated
choice. An important observation was that in certain problem classes
where small transition probabilities lead to local transfer matrices
close to unity, there is an inherent problem with the finite precision
of computer arithmetic if one wants to have fine Trotter
decompositions and to access comparatively large real times
simultaneously.


\ack

We wish to thank Malte Henkel, Andreas Kemper and Andreas
Schadschneider for useful discussions, AK and AS also for
communicating their results prior to publication and MH for a careful
reading of the manuscript. US is supported by a Gerhard-Hess prize of
the DFG.


\end{document}